\documentclass[journal]{IEEEtran}

\usepackage{cite}
\usepackage{graphicx}
\usepackage{amsfonts}
\usepackage{caption}
\usepackage{subcaption}
\usepackage{diagbox}

\usepackage{stfloats}
\usepackage{multirow}
\usepackage{amsmath}
\usepackage{algorithmic}
\usepackage{array}
\newcolumntype{C}[1]{>{\centering\let\newline\\\arraybackslash\hspace{0pt}}m{#1}}
\usepackage{url}
\usepackage{color,soul}
\usepackage{multirow}
\usepackage{hhline}

\usepackage{stfloats}
\usepackage{url}

\begin{document}
\title{Five Driving Forces of Multi-Access\\Edge Computing}



%
%

\author{ Madhusanka~Liyanage,~\IEEEmembership{Member,~IEEE,} Pawani~Porambage,~\IEEEmembership{Student Member,~IEEE,}
\\Aaron Yi Ding,~\IEEEmembership{Member,~IEEE,} 
\thanks{Madhusanka Liyanage and Pawani Porambage are with the Center for Wireless Communications, University of Oulu, Finland. e-mail:\{firstname.lastname\}@oulu.fi}
\thanks{Aaron Yi Ding is with the Department of Engineering Systems and Services, Delft University of Technology, Netherlands. e-mail: aaron.ding@tudelft.nl}
}




\maketitle

\begin{abstract}


The emergence of Multi-Access Edge Computing~(MEC) technology aims at extending cloud computing capabilities to the edge of the wireless access networks. MEC provides real-time, high-bandwidth, low-latency access to radio network resources, allowing operators to open their networks to a new ecosystem and value chain. Moreover, it will provide a new insight to the design of future 5th Generation (5G) wireless systems. This paper describes five key technologies, including Network Function Virtualization (NFV), Software Defined Networking (SDN), Network Slicing, Information Centric Networking (ICN) and Internet of Things (IoT), that intensify the widespread of MEC and its adoption. Our goal is to provide the associativity between MEC and these five driving technologies in 5G context while identifying the open challenges, future directions, and tangible integration paths.



\end{abstract}



\IEEEpeerreviewmaketitle
\vspace{-10pt}
\section{Introduction}
\label{sec:Introduction}

Multi-access edge computing~(MEC) is a relatively novel and an evolving networking paradigm that is currently standardized by the European Telecommunications Standards Institute (ETSI)\cite{hu2015mobile}. Its underlying principle is to extend cloud computing capabilities to the edge of cellular networks. Thereby, MEC is placing storage and computational resources from the Internet data centers all the way to the Radio Access Network~(RAN) edge where they are directly accessed by mobile devices and applications. 

Typically, MEC is characterized by key attributes such as closest proximity, ultra-low latency, location awareness, and network context information.  Firstly, since MEC servers are located close to the source of information, they have direct access to the devices and local resources. This proximity is very useful to capture key information for analytics and in Machine-to-Machine~(M2M) communication scenarios. Secondly, as the MEC servers are running close to end devices, they reduce the latency by reacting faster to improve user experience and to minimize congestion in other parts of the network. Thereby the user will experience ultra low latency and high bandwidth. Thirdly, when MEC servers are part of the wireless access network, a local service can leverage low-level signaling information to determine the location of each connected device. This property will encourage use cases which require location based services. Finally, MEC servers can exploit real-time network data such as radio conditions and network statistics to offer context related services to the particular applications. This can also differentiate the mobile broadband experience and be monetized. Thus, MEC plays a key role to achieve the vision of the fifth generation~(5G) wireless networks which are expected to reach 1~ms latency and high bandwidth along with the quantified users' quality of experience~(QoE).

In addition to  MEC, few other edge computing paradigms are also emerging such as Mobile Cloud Computing~(MCC), Fog computing, and Cloudlets. However, MEC is considered as better choice for 5G mobile networks than other owing to its compatibility with cellular networks and heavy backup by mobile standardization organizations such as ESTI.

In spite of its huge potential, the realization of MEC should be upheld by a multitude of underlying technologies. In this article, we examine five of those key enabling technologies, including Network Function Virtualization (NFV), Software Defined Networking (SDN), Information Centric Networking (ICN), Network Slicing, and Internet of Things (IoT), and illustrate how to utilize them to accelerate the adaption and development of MEC systems. Besides identifying open challenges, we also pinpoint tangible integration paths and future directions for MEC.

\vspace{-6pt}
\section{Network Function Virtualization}\label{NFV}

NFV proposes to utilize virtualization technologies to decouple physical network equipment from the functions that run on them \cite{mijumbi2016network}. Via this, different Virtual Network Functions (VNFs) can be implemented in software running on one or more industry standard physical servers. The VNFs can be relocated and instantiated at different physical network locations without necessarily requiring the purchase and installation of new hardware.  

\begin{figure}[htbp]
  \centering
  \includegraphics[width=0.5\textwidth]{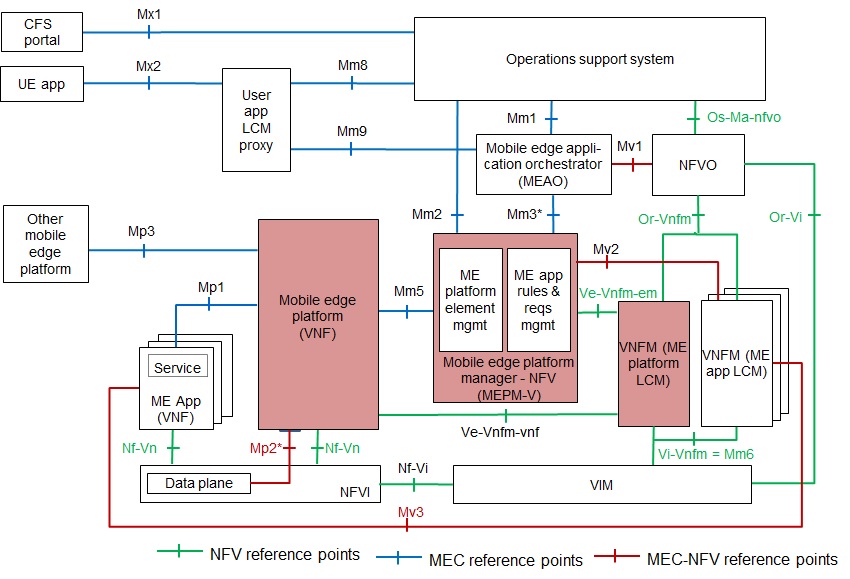}
  \caption{MEC in NFV Architecture\cite{MECinNFV}}
  \label{fig_MECNFV}
\end{figure}

Moreover, NFV is regarded as one of the key enablers for deployment of MEC\cite{gupta2016mobile} in 5G networks. Both NFV and MEC technologies can be used together in 5G mobile networks to elevate computing capacity to meet the increased networking demands. Both MEC and NFV share similar characteristics. For instance, MEC architecture is also based on a virtualized platform which is similar to NFV, as depicted in Figure \ref{fig_MECNFV}. Both technologies feature stackable components and each has a virtualization layer. Accordingly to ESTI\cite{hu2015mobile}, it is beneficial to reuse the infrastructure and infrastructure management of NFV to the largest extent possible, by hosting both VNFs and MEC applications on the same platform to enhance the computing experience. In addition, MEC can use the NFVI (NFV Infrastructure) as the virtualization platform to run mobile edge applications alongside other VNFs. Therefore, MEC applications also appear as VNFs in the NFV environment and parts of mobile edge orchestration can be delegated to the NFVO (NFV Orchestration)\cite{MECinNFV}.

On on hand, the main benefit of using MEC in NFV is to achieve low latency. MEC offers the possibility to host virtualized network functions closer to the user devices. Moreover, Service Function Chaining (SFC) is a part of NFV systems which use to connect to VNFs to follow an order and the data traffic to be flowed through the chain for carrying out the whole service. The use of MEC enhances the performance of SFC because deploying the functions closer to the end users decreases the latency and eliminates long-haul transmission of data traffic for carrying out the whole service.

On the other hand, the use of NFV in MEC will increase the scalability of MEC applications. NFV provides the high scalability by scaling in and out the network’s resources depending on need and application usage. Thus, combine use  of these two technologies will deliver a dynamic, quick and scalable computing platform for 5G ecosystem.

\vspace{-5pt}
\section{Software Defined Networking}\label{SDN}

SDN~\cite{kreutz2015software} is an emerging network concept that proposes to decouple the control plane functions from data plane of a switch. Moreover, it eliminates the use of vendor specific back-box hardware and promotes the use commodity  switches in data plane. 

Transferring network control functionalities to centralized entities demands SDN controllers to be located closer to data plane to reduce latency for critical applications. In this regard, MEC can be ideal solutions to satisfy the latency requirement. MEC complements the SDN advancement of the transformation of the mobile network into a softwarized networks, ensuring highly efficient network operation and service delivery\cite{blanco2017technology}. Thus, the popularity of SDN in various domains including 5G and IoT will further fuel the adaption of MEC concept as well.

\begin{figure}[htbp]
  \centering
  \includegraphics[width=0.4\textwidth]{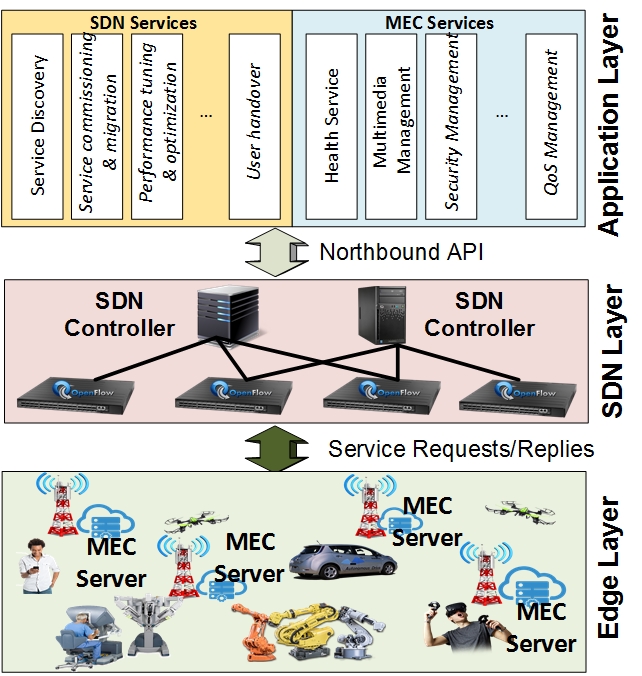}
  \caption{The use of SDN for MEC}
  \label{fig_MECSDN}
\end{figure}

SDN has capabilities of orchestrating the network, its services and  devices by hiding the complexities of the heterogeneous mobile environment from end users. Thus, SDN has a significant potential for mitigating the barriers and restrictions that multi-tier MEC infrastructure will encounter. Figure \ref{fig_MECSDN} illustrates a usage possibility of SDN for MEC ecosystem. For instance, The SDN  control mechanism can lower the complexity of MEC architecture by offering a novel approach to the networking and utilizing the available resources in a more efficient manner. SDN can dynamically route the traffic between tier-MEC servers and cloud servers  to provide the highest quality of service to end users. Moreover, SDN paradigm concentrates the network intelligence at the central software-based controller. This will relieve the relatively simpler MEC devices from executing the complex networking activities such as flow management, service discovery and orchestration. 

\vspace{-10pt}

\begin{table*}[b]
  \centering
        \caption{The role of MEC in different IoT  domains}
        \label{tab:mecIoTcharacteristics}
  \begin{tabular}{|p{1.5cm}|p{1.8cm}|p{1.2cm}|p{1.5cm}|p{10cm}|}
  \hline
      {\textbf{\centering{IoT Application}}}	
         &\textbf{{Data Capacity}}
         &\textbf{{Expected latency}}
         &\textbf{{Number of IoT Devices}}
       &\textbf{{Role of MEC}} \\ \hline \hline

Smart home      &  $\geq$ 10~MB
house per day    &      1~ms -1000~s &$\geq$10-100 per house & MEC offers reduced communication latency, easy instantiation and fast relocation. Moreover, MEC can process sensitive data locally by preserving the privacy. \\
\hline
Wearables  		      &   $<$ 1 GB per device per day     &     Several Hours&$\geq$1-10 per person & MEC allows to  deploy  storage, computing, and caching  in close proximity to satisfy wearable requirements such as scalability, short range and low power communication. \\ \hline
\hline
Smart city         & $\geq$100-1000 TB per city per day       &$\leq$1ms& Few millions per city& Data can be processed at the edge of the network to provide low latency, location awareness and scalability\\
\hline
Retail  &  100~Mbps - 1~Gbps  & $\leq$1~ms & $\geq$100-1000 per shop & On-site MEC servers can locally process huge volumes of data generated by different IoT systems such intelligent payment solutions, facial recognition systems, smart vending machines. \\
\hline
Smart energy	       &  $\geq$  100,000 GB per day   & 1ms - 10 mins & $\geq$ 1 billion per grid & MEC allows  the computation to be performed closer to the data source by reducing impacts of bandwidth bottlenecks and communication delays due to poor network connectivity and huge data generation. Moreover, MEC increase the security and reduce the attack propagation by  enforcing security mechanisms closer to the end devices.  \\
\hline 
Remote surgery    	    & $\geq$1~GBps      & 1 - 200~ms &$\geq$10-100 per surgery& 
Ensure the ultra-low latency and uninterrupted communication for remote areas\\
\hline

Autonomous vehicles	      &  $\geq$ 100 GB per vehicle per day  &$\leq$1~ms&50-200 per vehicle &  MEC can improve the operational functions such as  real-time traffic monitoring,  continuous sensing in vehicles,  Infotainment applications and security~ by fulfilling the latency, reliability, fast big data  processing, and throughput requirements\\
\hline
AR/VR     & $\geq$1~GBps      & $\leq$1~ms &$\geq$0.2 million globally &  Migrating computationally intensive tasks to edge servers will increase the computational capacity of VR devices and save their battery-life. Furthermore, MEC platforms offer scalability for by enabling high capacity and low latency wireless coverage for large venues like stadiums or smart cities with a massive density of users to enjoy the AR and VR experience.\\
\hline

Gaming	     &$\geq$10~Mbps     & $\leq$10~ms& $\geq$1 billion globally & 
MEC can improve user experience for delay-sensitive game users by offloading the resource-intensive applications to the edge servers that are located in the nearest proximity.  
\\
\hline
Industrial IoT (IIoT)   & $\geq$ 100,000 GB per day     & $\leq$1~ms & $\geq$ 1 million per factory & MEC enabling future IIoT applications by addressing the shortcomings of M2M communication (e.g. latency, resilience, cost, peer-to-peer, connectivity, security). Moreover, real-time edge analytics and enhanced edge security  help to create new IIoT services.\\ \hline  \hline

Weather Monitoring     &    Few MBs per station  per day     & Minutes to hours & 5-10 per station & MEC process the information closer to sensor and removes the burden of sending raw data over a network with limited bandwidth\\ \hline
Farming and Poultry        &   $\geq$ 1 GB per farm        & Several hours & 5 -100,000 per farm    & On-site MEC servers can analyze collected big data without real-time uploading to a remote cloud. Thus, MEC can directly reduce the overhead on data access, synchronization and storage.\\ \hline



  \hline
  \end{tabular}
  \end{table*}

\vspace{-10pt}
\section{Information Centric Networking}\label{ICN}

To support the ever increasing bandwidth demand and low latency for Internet applications such as 4K/HD Videos, 3D games,  and AR (Augmented Reality)/ VR (Virtual Reality), several networking technologies are developed over the past decades. Most of these technologies focused on utilizing caching, replication and content distribution in optimum ways. Similar to the MEC, ICN is another network concepts which can satisfy this demand\cite{vasilakos2015information}. In particular, ICN is an Internet architecture that puts information at the center where it needs to be and replaces the client-server model by proposing  a new publish-subscribe model.



Several benefits can be achieved by exploiting the synergy between MEC and ICN. The use of ICN can solve some MEC issues related to the content delivery and application level reconfiguration. ICN can offer high speed content delivery between the MEC and central cloud systems. Application level reconfiguration is challenging in MEC systems, since a session re-initialization is required whenever a session is being served by a non-optimal service instance. This process always increases the  session migration delay and significantly affects the low latency applications. ICN can reduce the application level reconfiguration delay by minimizing the network configuration delay for MEC applications. Due to the service-centric networking characteristics in ICN technology, it allows fast resolution of named service instances\cite{ravindran2017realizing}. 

The coexistence of ICN and MEC can also improve the performance of the edge storage and caching function at the edge networks. This is enabled by two features of ICN naming location independent data replication and opportunistic caching at strategic points in the network. These features benefit both realtime and non-realtime 5G applications where a set of users share the same content\cite{ravindran2017realizing}. 

ICN will significantly improve the efficiency of session mobility in MEC networks with the optimal operational cost and bandwidth utilization for signaling traffic. In contrast to the anchor-based mobility approach used in current MEC networks, ICN handles session mobility by using application bound identifier and location split principles which have significantly reduced control and user plane overheads. 
  
The real-time context aware applications could be accomplished with the correct coordination between MEC platforms. ICN provides considerable opportunities for context-aware data distribution in the networks by allowing content distribution over unreliable radio links and transparent mobility between heterogeneous network. Due the latency support of MEC, ICN-MEC integration is important to provide services for high mobility 5G applications such as tactile Internet and autonomous vehicles.


\begin{figure}[ht]
  \centering
  \includegraphics[width=0.50\textwidth]{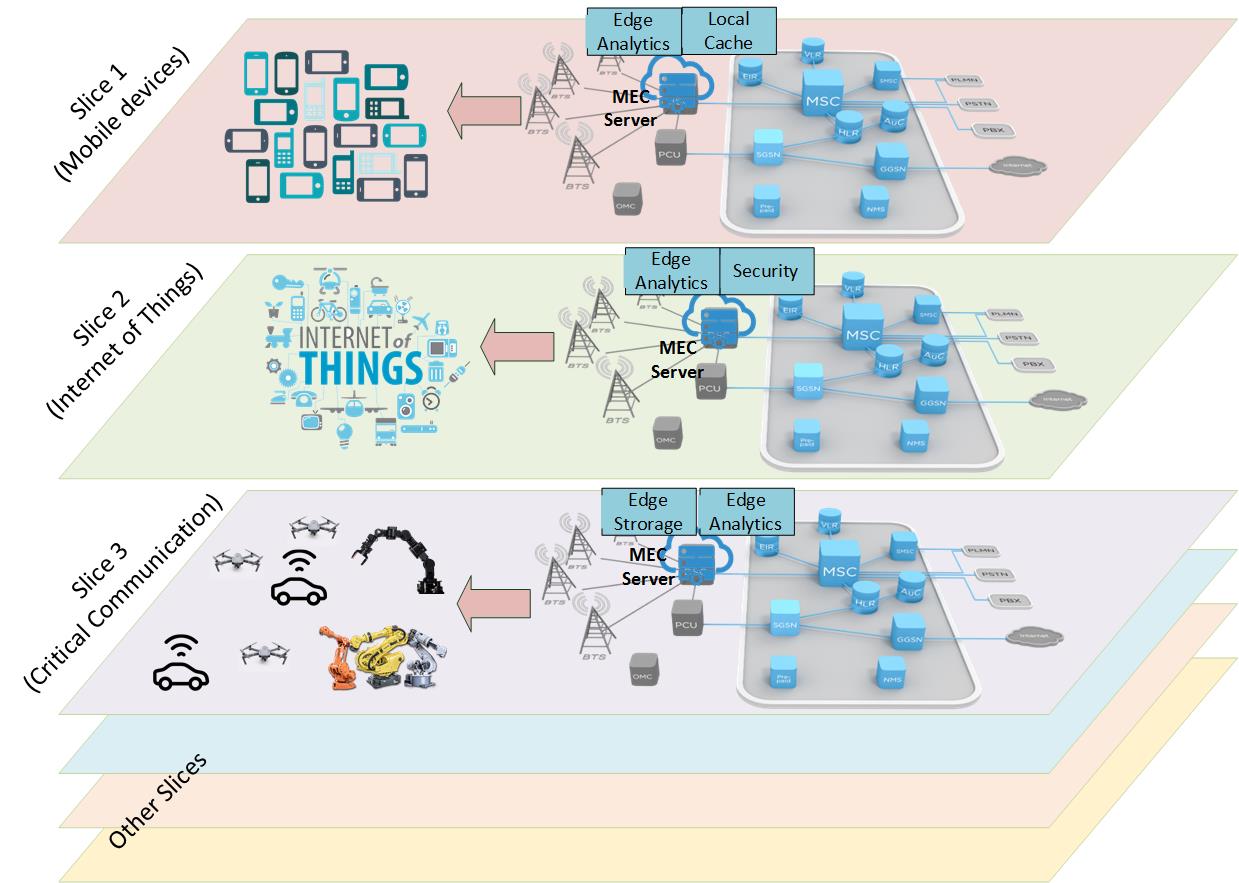}
  \caption{Use of Network Slicing and MEC in different 5G applications}
  \label{fig_SlicingUC}
\end{figure}

\vspace{-5pt}
\section{Network Slicing}\label{NS}


Network Slicing has also emerged as a key concept for providing an agile and dynamic networking platform on demand. It allows multiple virtual networks to be created on top of a common shared physical infrastructure\cite{samdanis2016network}. 

 
Upcoming 5G architecture  will utilize both MEC and network slicing\cite{alliance2016description} along with other technologies.  MEC and network slicing can be used together to provide in different application domains in 5G. One of such application domains is Massive IoT\cite{nikaein2015network}. In order to support massive IoT systems, the network should be able to satisfy the requirements such as massive cost reduction, network scalability and edge analytics. Network slicing with the MEC-based analytics and security assets can be used to deliver these requirements. Another use case is critical communications for delay critical applications such Tactile Internet, autonomous driving and industrial Internet.   The key requirements to enable the critical communications are  reduced latency and traffic prioritization. While MEC can be used to reduce the latency, network slicing can support traffic prioritization. As shown in Figure \ref{fig_SlicingUC}, network slicing can help to divide the MEC resources to different slices based on tenants' demands. 

Moreover, network slicing can enable dynamic and in short life cycles for network services. This feature will enable new value creation opportunities where resource sharing among virtual MNOs (Mobile Network Operators), services and applications in time share manner.  MEC provides edge analytics and faster security assets for better network slicing which will lead to massive cost reduction and increase of network scalability.

MEC has been identified as one of the key attributes to realize the aforementioned network slicing extensions in 3GPP (3rd Generation Partnership Project) toward full multi-tenancy. Thus, the synergy between MEC and network slicing will play a critical role in  the deployment of 5G  applications.

\section{Internet of Things (IoT)}

IoT is an ecosystem of connected physical objects that are accessible through the Internet. Currently, IoT supports a myriads of application areas. Since IoT devices are low powered with low memory footprints and less processing power, centralized cloud computing is used for storing and processing  in many IoT environments\cite{porambage2018survey}.



Although cloud computing enables the outsourcing of storage and processing functionalities, a conventional cloud faces several challenges such as the single point of failure, lack of location awareness, reachability, and high wide area network (WAN) latency. 
In this content, MEC  has risen up in order to fill the gap between the centralized cloud and IoT devices by providing many mutual advantages\cite{bonomi2014fog, porambage2018survey}. To illustrate the connection, we explain the role of MEC in different IoT domains in Table \ref{tab:mecIoTcharacteristics}.

 Firstly, MEC allows IoT to filter large volumes of data generated by massive IoT applications at the edge of the network, saving the time and cost of transmitting data to the data centers.
 Moreover, if the IoT devices do not need global level services, with the cloud computing capability, MEC servers can process the packets and provide the required services without sending to the core network.  This will reduce the traffic volume in the core network and reduce the latency. 

Secondly, MEC facilitates rapid decision making based on the locally processed data by reducing the End-to-End~(E2E) delay. This is very important in the critical IoT applications (e.g., remote surgeries, smart grid, autonomous vehicles, and video conferencing), which have very high demands of reliability, availability, and low latency.

Thirdly, MEC will improve the scalability as the number of IoT connections increase and reduce the drain on the battery due to the less transmission time between the device and application server.

\vspace{-10pt}
\section{Challenges, Future Directions and Integration Path}

Up to this point, we illustrate the mutual benefits of using each driving technology in MEC. Table \ref{tab:mecOthercharacteristics} summarizes how can we utilize them to improve various features and create the demands (e.g., IoT deployment) for MEC. 

\begin{table}[htbp]
  \centering
        \caption{The impact of driving technologies to enhance MEC}
        \label{tab:mecOthercharacteristics}
  \begin{tabular}{|p{4cm}|p{0.4cm}|p{0.4cm}|p{0.4cm}|p{0.4cm}|p{0.4cm}|}
  \hline
      {\textbf{\centering{MEC Features}}}	
         &\textbf{{NFV}}
         &\textbf{{SDN}}
         &\textbf{{ICN}}
       &\textbf{{NS}}
       &\textbf{{IoT}}\\ \hline \hline

\textbf{Support for Low Latency}     &      &   \checkmark    & \checkmark & &  \\ \hline
\textbf{Resource Optimization}        & \checkmark      & \checkmark &  & \checkmark & \\ \hline
\textbf{Dynamic Resource Allocation}        & \checkmark      &  &  &  \checkmark & \\ \hline
\textbf{Support for Edge Caching}       &        & & \checkmark & & \checkmark \\ \hline
\textbf{Increased Security}       & \checkmark    & & & \checkmark & \\ \hline
\textbf{Increased Privacy}       &     & & & \checkmark & \\ \hline
\textbf{Increased Scalability}       & \checkmark       & \checkmark & \checkmark & \checkmark & \checkmark \\ \hline
\textbf{Reduce the Operational Cost}       & \checkmark    & \checkmark & \checkmark & \checkmark & \\ \hline
\textbf{Increase Flexibility}       & \checkmark    & \checkmark & & \checkmark &\\ \hline
\textbf{Increase Orchestration}       & \checkmark    & \checkmark &  & & \\ \hline
\textbf{Dynamic Routing and Traffic Optimization }       &    & \checkmark & \checkmark  & & \\ \hline
\textbf{Support for Fast Mobility}       &    &  & \checkmark  & \checkmark & \\ \hline
\textbf{Service Diversity}       & \checkmark   &  & \checkmark  & & \checkmark\\ \hline
\end{tabular}
  \end{table}
  
In this section we discuss the obstacles and challenges related to the integration of different technologies. Based on each category, we share our insights on future directions and integration plan.
\vspace{-10pt}
\subsection{NFV}
Main research challenges and obstacles for NFV-MEC integration are the absent of standards, immaturity of technologies, deployment complexity and new security risks. These obstacles and challenges to be address to achieve the full integration benefits of MEC and NFV in 5G networks.

Both NFV and MEC are recent technology are evolving through the phases of implementation and requires standardization emanating from collaboration of industry and researchers over an agreed platform. Specially, the interfaces and architectural components which are required to MEC NFV integration should be defined at global level. Otherwise,  wide spread adaptation will hinder due to the compatibility issues.

Most NFV and MEC projects will face a steep learning curve in getting their infrastructure to work as expected because of their heavy dependency on non-standardized implementations. Due to the immaturity of both technologies, the updates are released frequently. Thus,  keeping up in an operational deployment model is hard to achieve.

Minimizing the latencies through optimal utilization of resources can be achieved with the efficient deployment of MEC services. However, it is difficult to optimize the MEC services when they are depending on complex system components such as NFV. Specially, de-facto NFV standard implementations such as OpenStack, is also notoriously difficult to learn, deploy, and use. 


NFV-MEC integration creates several new security challenges as well. On one hand, MEC introduces software components such as MEPM, VNFM to NFV deployments. These components are not a part of traditional NFV  model and they cause to create a “longer chain of trust”. On the other hand, the NFV features such as  resource pooling can lead to the sharing of risk between multiple unrelated MEC domain. For instance, an attack on a certain VNF might affect other VNFs running on the same VM (Virtual Machine) or physical server. 
\vspace{-10pt}
\subsection{SDN}

There are a few research challenges such as new security risks and modification of APIs (Application Programmable Interfaces) which need to be addressed to achieve the full integration benefits of MEC and SDN technologies.

The use of SDN introduces new security threats on MEC systems. SDN has many security threats including SDN protocol weaknesses,  information disclosure through interception, flow poisoning, side channel attacks and DoS (Denial of Service) on SDN controller. Since MEC extends cloud computing capabilities to the edge of mobile networks, the level of protection that can be offered to the edge hosts is low compared to what is obtainable in traditional large data centers. Thus, the integration of MEC with SDN will further reduce the protection of SDN systems and the impact of SDN based attacks will results the service degradation on MEC systems.

In addition, MEC has its own security issues. For instance, proposed architectural modifications in MEC create a number of security vulnerabilities such as malicious mode problems, privacy leakages, and VM manipulation. The impact of these MEC threats on the open network based SDN is more devastating. In contrast to traditional black-box type network devices, SDN uses software programmable common standard backhaul devices. It will not only ease the work of network administrators but also allow malicious attackers to deploy attacks. 

MEC-SDN interworking will also introduce several connectivity challenges. Similar to SDN southbound, northbound, east/west interfaces, MEC also have three interfaces, namely 1) Northbound connections, which are the connections between MEC servers  and a Cloud service (public or private),  2) Southbound connections, which are the connections between the MEC servers and the Edge devices and 3) East/West connections, which are the connections between MEC themselves, so that MEC servers can communicate without requiring, cloud connectivity. Since they serve similar purpose in high level, it is necessary to merge similar interface to reduce the signaling overheard. Furthermore, use of many inferences  reduces the security of the network.

It is also necessary to define APls so that applications and services can program network functions and SDN network directly bypassing control and management to optimize the performance, For instance, such APIs are need to support ultra-low latency applications. Otherwise, information exchange between MEC and SDN systems will introduce additional delays in network operations.

\subsection{ICN}


As per today, MEC and ICN are complementary concepts\cite{grewe2017information} that are mostly deployed independently. Good coordination is needed to obtain the best outcomes of their synergies. Proper APIs have to be defined in order to communicate between the systems. Although there is on-going research to define interfaces for MEC-NFV and MEC-SDN integrations, the interface for MEC-ICN communication is yet to be defined. Moreover, ICN-MEC integration requires a cross-domain industry collaborations and standard development organizations.

Specially, it is important to develop system control orchestrator and/or coordination architecture to enable the cooperation of two systems. Moreover, such architecture should focused on automatic and autonomic system control rather than the traditional provisioning/configuration or distributed control of networking systems.

The real advantages of MEC can be achieved by obtaining context information such as users' location, other users in vicinity, condition  and resources in the environment. Although ICN can provide different levels of context information (application, network and device level), their simultaneous retrievals are still challenging. Most of the current ICN research is focused on providing the basic functionality, rather than on utilizing the available context information to improve network parameters such as Quality of Service (QoS) and Quality of Experience (QoE). In ICN-MEC collaborations, it is required to examine typical scenarios encompassing different 5G applications (AR/VR, autonomous driving, Tactile Internet) with varying context.

Moreover, MEC-ICN integrated systems face severe challenges due to  authorization and access control issues in ICN systems. User/level authorization is a significant challenge in ICN due to the the lack of user-to-server authentication. ICN systems can not use traditional access control schemes based on Access Control List (ACL). The in-network caching function of ICN enable the possibility to deliver Named Data Objects (NDOs) on demand basis. In this environment, ICN entities have to maintain an identical control policy over NDOs for each consumer to support ACL based access control. However, such mechanism is challenging due to privacy issues and computational overhead.

\vspace{-10pt}
\subsection{Network Slicing}

There are several challenges that has to be address to achieve the true cooperate benefits of network slicing and MEC technologies.

The inter-system vertical coordination between the two technologies should be clearly structured and modeled for efficient information sharing. This vertical coordination can be achieved via two ways. First method is to define APIs between management systems of slicing and MEC to share the available resources. Second method is to use physical resource coordination aimed to efficient resource handling through policy and analytics. However, standardization these interfaces have to be performed to synchronize the various research and development activities around the globe.

If the MEC servers can offer service composition with fine-grained network functions, it will enhance the scalability to support different vendors. Each coarse grained function at the MEC server can be further divided into many sub-functions. Nevertheless, the granularity of those networking functions should be defined carefully in such a way to comply with the available standardized interfaces. Otherwise, same as in the above case the interfaces need to be recognized by the respective standardization entities. 

When multiple RATs are accommodating the 5G paradigms, there should be some ways to multiplex them on specialized or dedicated hardware. Although network slicing may lead to virtualize RAN instances, it is mandate to ensure radio resource isolation and manage efficiency. In order to assist RAN virtualization for slicing, Software Defined ) RAN controllers can be run at the MEC servers.

Even though the high-level description of a concrete slice in terms of infrastructure and network functions is defined, the physical realization of E2E slice orchestration is still to be established. 
As the intermediary edge computing platforms, the MEC servers play a vital role to support E2E slice orchestration and management by correlating cloud and radio resources.

\vspace{-10pt}
\subsection{IoT}
Although IoT is becoming a mature technology, the constant addition of new IoT devices to the consumer market and their versatile connectivities create myriads of security and scalability issues. When MEC is coincided with IoT, these challenges will arise in a broader manner. Security, privacy, and trust management are three important research areas of IoT which have many synergies. Edge computing users are increasingly vulnerable to security threats as more IoT devices and applications use the edge to transmit information. The heightened exposure of user's data in MEC and IoT may create many possible ways which the sensitive data can be breached. With the growing intelligence of the smart devices, there can be the possibility of one IoT device betraying the implicit trust of another IoT device. Typically, the IoT devices are programmed to automatically trust another connected device and to share data without a validation process. If all the devices natively trust each other and share data, it is hard to identify whether there is a misbehaving device. This may create a big issue specially with the absence of a perimeter around the network edge prohibits firewalls to block out MEC security threats. Moreover, the frequent back and forth transfer of data from a device to the network edge increases the opportunity for breaching data and hacking devices. In MEC systems, it is challenging to identify, authenticate, and authorize devices and the data they generate from the edge to cloud and back while maintaining a latency of some milliseconds.

Scalability and mobility are two other important features that MEC provides for IoT applications. In many applications, IoT devices will require scalability of services by applying load balancing mechanisms which can be managed by a cloud orchestrator in MEC nodes. As edge devices uses different access technologies including 3G, 4G, 5G, Wi-Fi and Wi-Max the aspect of heterogeneity and scalability should be catered in smooth functioning of MEC operations. Similarly, it is challenging to achieve non-negligible impact on caching and computation offloading decisions with the user mobility which will cause frequent handovers among edge servers.






\subsection{Integration Path}

We explain here the integration paths and pinpoint tangible steps to realize them.

\begin{figure}[b]
  \centering
  \includegraphics[width=0.45\textwidth]{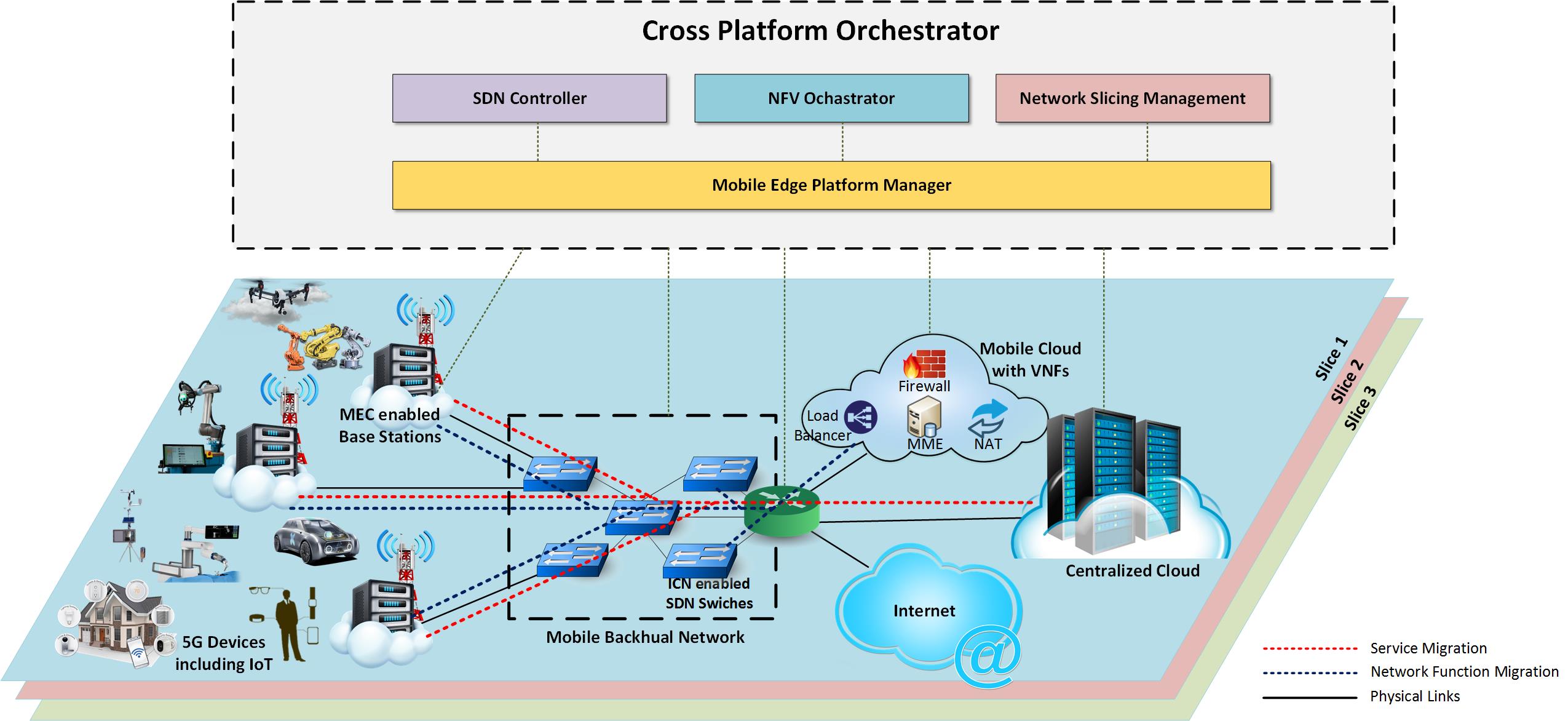}
  \caption{Integration of Driving Technologies in MEC Systems}
  \label{fig_integrated}
\end{figure}

\subsubsection{Control Level Orchestration}  To achieve added benefits in MEC-enabled networks, different technologies must work together. Given that 5G is going to integrate those five driving technologies with MEC, as depicted in Figure \ref{fig_integrated}, it is expected to enable them simultaneously in 5G. Meanwhile, such integration will face challenges at control level. For instance, SDN, NFV, Network Slicing and  MEC are independent technologies. Each technology utilizes its own orchestrator and management entities such as SDN controller, NFV orchestrator, network slicing manager and mobile edge platform manager. 
In this respect, a synergy between these control entities is needed to jointly optimize the network resources and also to create efficient  Service Function Chains (SFCs) for each user application.  

\subsubsection{Synchronization of Standardization Process}
In order to achieve orchestration in MEC systems, different technological components need to inter-communicate with each other. This demands that communication interfaces have to be defined at architecture level. However, the standardization of different technologies are coordinated by different organizations, e.g., MEC and NFV by ESTI, SDN by ONF, ICN by IETF, IoT by IEEE,and Open Internet Consortium (OIC).
Therefore, the cooperative efforts of different standardization bodies are needed to make synergies between the interfaces defined for different technologies. As a good example, ETSI has already started defining the interfaces for NFV and MEC integration (Figure \ref{fig_MECNFV}). However, these efforts need to expand for other technology domains as well.

\subsubsection{Hardware Limitations and Dependencies}
The integration of driving technologies demands changes not only at the control plane but also at the data plane and hw/sw components. For instance, SDN-enabled switches and devices are needed at the infrastructure layer to implement SDN. Similarly, ICN enabled switches are needed to enable ICN functionality. Installation of such multi-technology hardware is challenging. First, standardization of different technology integration should be finalized so that vendors can start building such multi-technology hardware equipments. Second, extensive hardware resources are needed to implement multi-technology concepts. 
Therefore, these hardware limitations and dependencies has to be solve to obtain full benefits of  technology integration. 


\subsubsection{Security and Privacy} 

To deploy MEC at large scale, security and privacy must be enforced by all five technologies across multiple layers in MEC. In particular, since IoT is closely coupled with public and enterprise infrastructure, this cyber-physical setting renders ample use cases and also generates security challenges for MEC to integrate NFV, SDN, ICN and network slicing. The principle of security and privacy by design should hence be mandated in the integration process. 

\subsubsection{AI as a key integration enabler}
Recently artificial intelligence (AI) and machine learning (ML) have 
been resorted to create smarter and autonomous wireless systems\cite{iliadis2018artificial}.
In the 5G context, AI can directly benefit the driving technologies such as SDN and NFV to be integrated into MEC. For instance, AI-based edge orchestrators can be used for better system and host level management functions for various NFV based use cases.
AI and MEC together (i.e., edge automation) will combat towards low latency for real-time services, better orchestration, enhanced security, and backhaul cost savings.



\vspace{-10pt}

\section{Conclusion}

This paper analyzes five integration technological directions that can accelerate the utilization and deployment of MEC  in 5G mobile networks, including NFV, SDN, ICN, Network Slicing and IoT. We highlight the benefits of using each technology in MEC systems as well as identify the remaining challenges and integration path to realize the full integration benefits. With careful integration, the suite of these solutions will form the future of MEC deployments.



%

\vspace{-10pt}
\section*{Acknowledgment}
This work has been performed under the framework of  6Genesis Flagship (grant 318927) which is funded by the Academy of Finland.

\vspace{-10pt}
\bibliographystyle{IEEEtran} 
\bibliography{bibtex/bib/MECIoTSurvey}

\vspace{-20pt}
\begin{IEEEbiography}[{\includegraphics[width=1in,height=1.25in,clip,keepaspectratio]{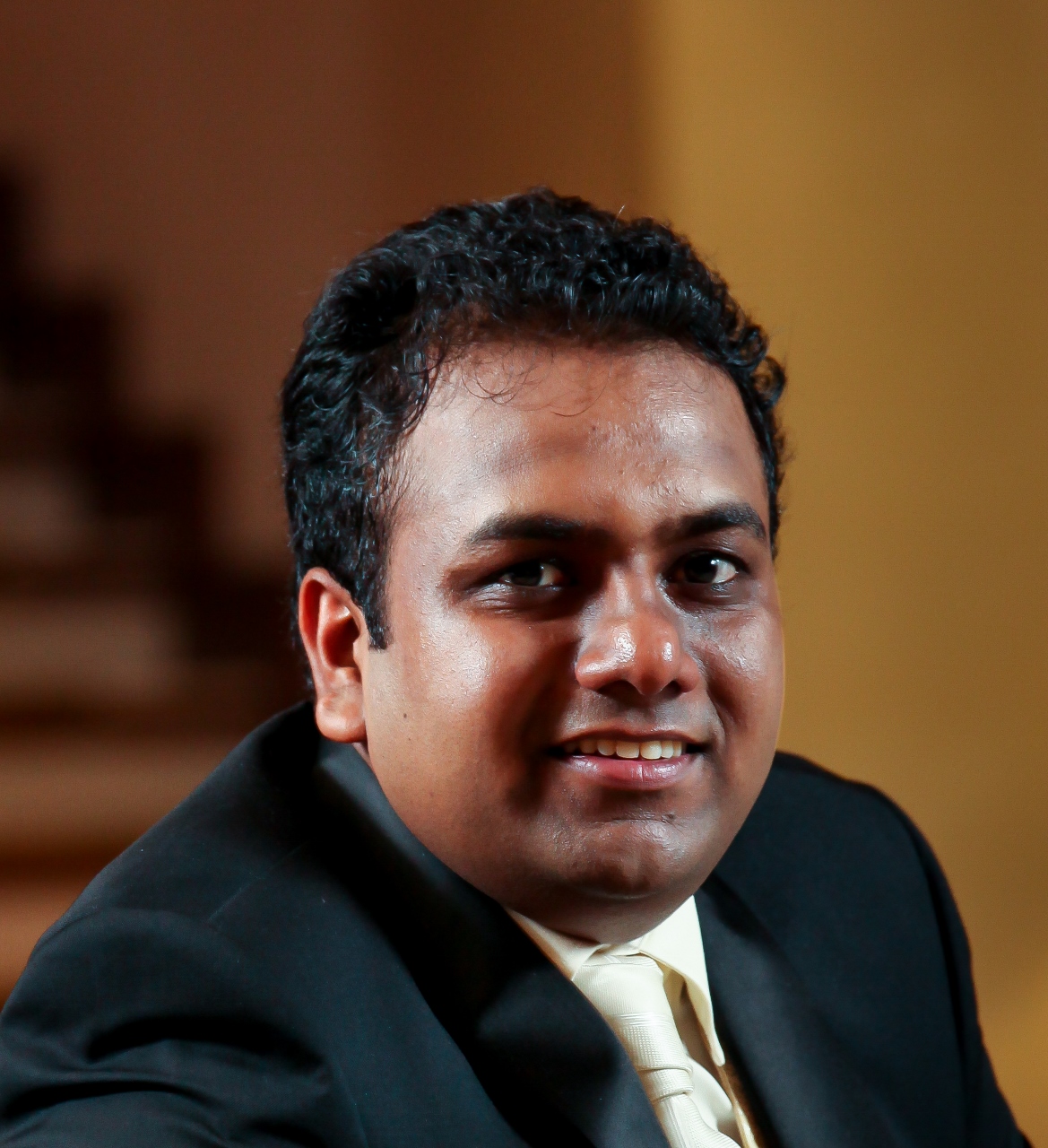}}]{Madhusanka Liyanage}
received a Doctor of Science (DSc.) degree in communication engineering from the University of Oulu, Oulu, Finland. He is currently an adjunct professor at the Centre for Wireless Communications, University of Oulu, Oulu, Finland.  He has received several best paper awards and co-authored more than 40 scientific publications including two edited books with Wiley. His research interests are SDN, IoT, Blockchain, MEC, mobile and virtual network security. Contact him at madhusanka.liyanage@oulu.fi 
\end{IEEEbiography}
\vspace{-20pt}
\begin{IEEEbiography}[{\includegraphics[width=1in,height=1.25in,clip,keepaspectratio]{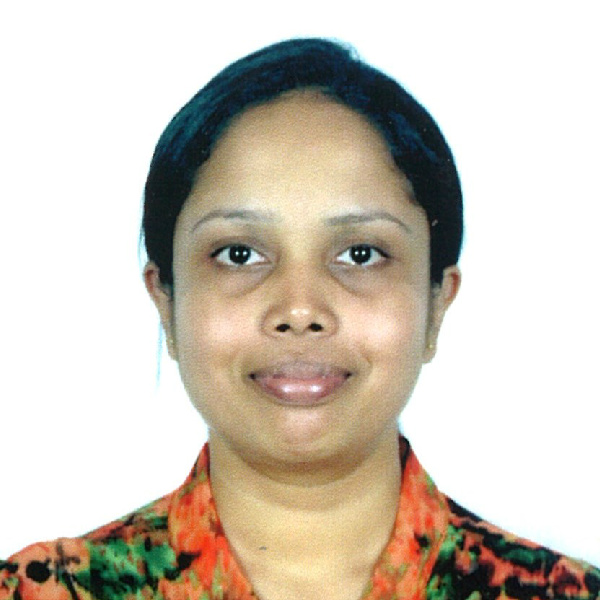}}]{Pawani Porambage}
Pawani Porambage is postdoctoral researcher at the Centre for Wireless Communications, University of Oulu, Finland. She obtained the Doctor of Science (DSc.) degree in  from the University of Oulu, Oulu, Finland. She also received her Bachelor Degree in Electronics and Telecommunication Engineering in 2010 from University of Moratuwa, Sri Lanka and her Master’s Degree in Ubiquitous Networking and Computer Networking in 2012 from University of Nice Sophia-Anipolis, France. Her main research interests include lightweight security protocols, security and privacy on IoT and MEC, and Wireless Sensor Networks.
\end{IEEEbiography}
\vspace{-20pt}
\begin{IEEEbiography}[{\includegraphics[width=1in,height=1.25in,clip,keepaspectratio]{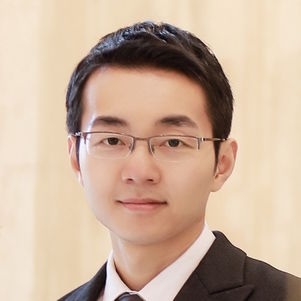}}]{Aaron Yi Ding}
is tenure-track Assistant Professor at TU Delft. He has over 11 years of R\&D experience in EU, UK and USA, working on edge computing, lightweight virtualization and Internet of Things (IoT). He obtained his MSc and PhD both from the University of Helsinki. Prior to TU Delft, he worked at TU Munich (2016-2018) and University of Helsinki (2007-2016). He conducted research at Columbia University (2014) and at University of Cambridge (2013), advised by Prof. Henning Schulzrinne and Prof. Jon Crowcroft, respectively.
\end{IEEEbiography}

\end{document}